\documentstyle[preprint,aps]{revtex}
\input epsf
\begin{document}
\draft
\title{Anomaly in Numerical Integrations of the
KPZ Equation and Improved Discretization}
\author{Chi-Hang Lam and F.G. Shin}
\address{
Department of Applied Physics, Hong Kong Polytechnic University, Hung Hom, Hong Kong
}
\date{\today}
\maketitle
\begin{abstract}
We demonstrate and explain that conventional finite difference schemes
for direct numerical integration do $not$ approximate the continuum
Kardar-Parisi-Zhang (KPZ) equation due to microscopic roughness. The
effective diffusion coefficient is found to be inconsistent with the
nominal one. We propose a novel discretization in 1+1 dimensions which
does not suffer from this deficiency and elucidates the reliability and
limitations of direct integration approaches.
\end{abstract}

The Kardar-Parisi-Zhang (KPZ) equation has been very successful in
describing a class of dynamical self-affine interfaces
\cite{Review}. Numerous simulations on discrete models for vapor
deposition, bacterial colony growth, directed polymers, etc. show
agreements with KPZ predictions. Being the simplest nonlinear
stochastic evolution equation for interfaces, the KPZ equation is
believed to be relevant to a large diversity of phenomena although
experimental verifications has been controversial \cite{Review}. Many
numerical investigations on the subject have concentrated on discrete
models. This work focuses on another important approach, namely,
direct numerical integration of the KPZ equation. Amar
and Family first conducted such large-scale integrations
\cite{Amar}. They found scaling exponents of the resulting interfaces
in agreement with those from discrete models. This conclusion is
supported subsequently by more accurate works indicating the validity
of the KPZ approach
\cite{Moser91,Moser94}. 

However, it has been observed that the discretized equations in the
numerical integration of the KPZ equation admit peculiar properties
not fully compatible with their continuum counterparts
\cite{Moser91,Amar93,Newman}. By applying Lam and Sander's inverse
method \cite{Lam}, we will give a quantitative demonstration and
theoretical explanation of an abnormal behavior of the diffusion
coefficient. We propose a novel discretization for the numerical
integration of KPZ interfaces in 1+1 dimensions. Our discrete
equations behaves in a much more predictable way as proved by exact
solution of its steady state properties. The results should clarify the
reliability and limitations of conventional numerical integration
techniques on the KPZ equation.  In addition, conventional direct
numerical integration schemes for the KPZ equation are inefficient and
numerically rather unstable at high nonlinearity
\cite{Amar,Moser91,Moser94}. We will give a
quantitative evaluation of this instability. In contrast, the new
discrete equations can be integrated substantially more efficiently
with much improved stability. 

The KPZ equation gives the local rate of growth of the coarse-grained
height profile $h(x,t)$ of an interface at substrate position $x$ and
time $t$
\cite{Review}:
\begin{equation}
\label{KPZ}
\frac{\partial h}{\partial t} = c +  
\nu\nabla^2 h+\frac{\lambda}{2}( \nabla h)^2  
 +\eta(x,t) ,
\end{equation} 
where $c$, $\nu$ and $\lambda$ are the average growth rate, the
diffusion coefficient and the nonlinear parameter respectively.  There
is an implicit lower wavelength cutoff below which $h$ is smooth.  The
noise $\eta$ has a Gaussian distribution and mean $0$ and a correlator
$<\eta(x,t)\eta(x',t')> = 2 D \delta( x-x') \delta(t-t') $.  Most previous
works on the numerical integration of the KPZ equation adopt the finite
difference and Euler's method with the
following equation \cite{Amar,Moser91,Moser94}:
\begin{eqnarray} 
\label{org_dis}
h^{n+1}_i&=&h^n_{i}+\Delta t[\nu_0(h^n_{i+1}+ h^n_{i-1} -2 h^n_{i}) 
\nonumber\\& & 
+ ({\lambda_0}/{8}) ( {h^n_{i+1} -h^n_{i-1}} )^2 ]
 + \sqrt{2D_0\Delta t} \xi^n_i ,
\end{eqnarray}
where $h^{n}_i$ approximates $h(i\Delta x,n\Delta t)$ with periodic
boundary conditions and every $\xi^n_i$ is an independent Gaussian
variable with zero mean and unit variance. The subscripted parameters
$\nu_0$, $\lambda_0$ and $D_0$ are nominal values used in the
iteration to be distinguished from the continuum values in the KPZ
description in Eq. (\ref{KPZ}).  Following previous works
\cite{Amar,Moser91,Review}, the spatial
step size $\Delta x$ is taken to be 1. Other choices will be discussed
later. The implicit lower wavelength cutoff here is effectively 1 due
to the spatial discretization. The temporal step size $\Delta t$ is
taken to be small enough so that decreasing its value further will not
alter the results.

In fact, it can be verified easily that as long as it is numerically
stable, $\Delta t$ need not be small before the KPZ scaling exponents
can be computed accurately. This is because the discrete equation with
finite $\Delta t$ is by itself in the KPZ universality class similar
to many discrete models due to symmetry considerations. The reason for
taking a small $\Delta t$ is to allow the discrete equation to
approximate the continuum KPZ equation. However, we suggest that
finite differencing is not a good approximation because of microscopic
roughness, although this does not alter the
scaling exponents due to universality.

\begin{figure}
\epsfysize 8in \epsfbox{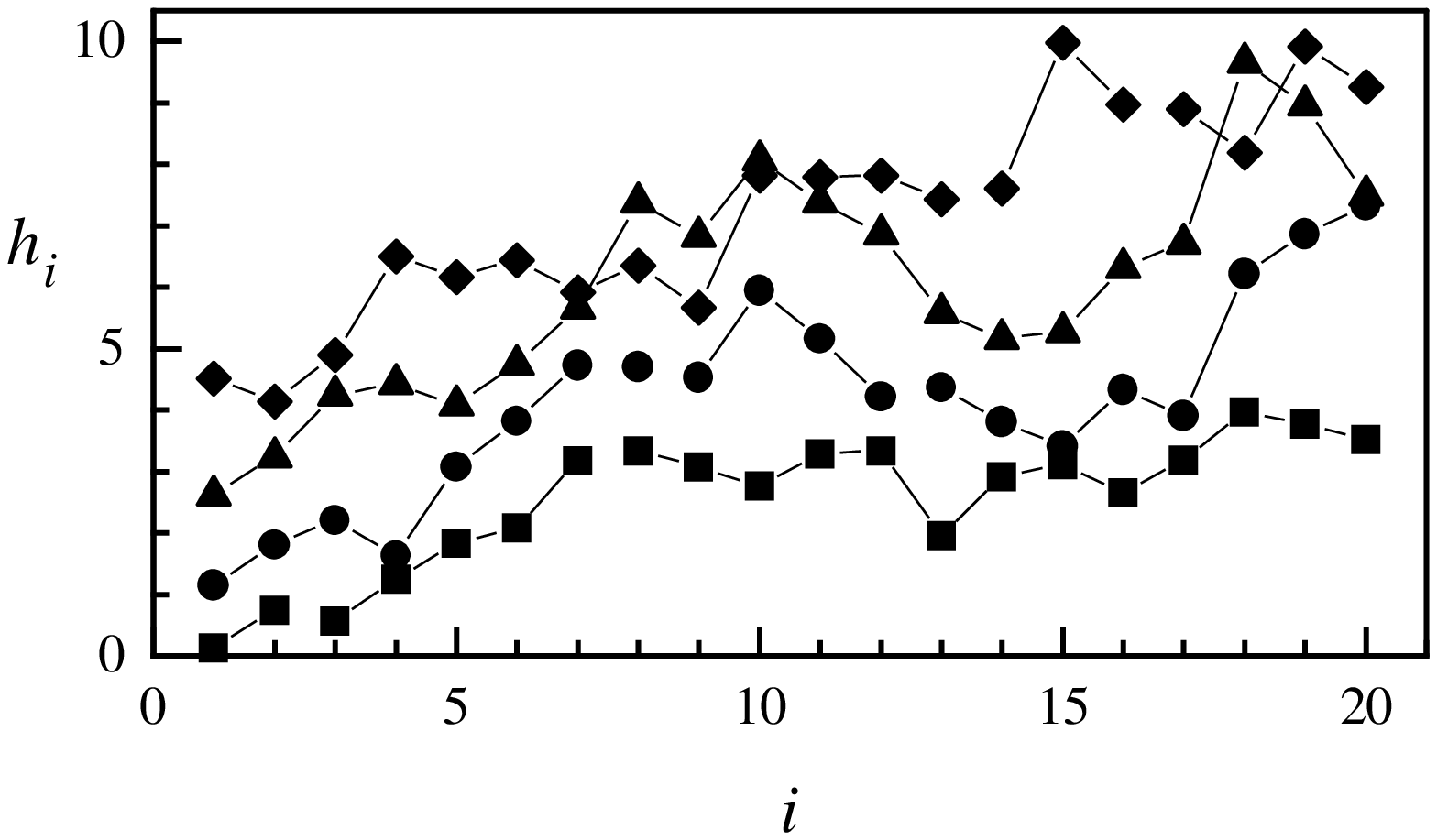}
\caption{Snapshots of a segment of an interface
generated by numerical integration. The time between two consecutive
snapshots is $0.2$ corresponding to 20 iterations.
\label{Finterface}}
\end{figure}

Naively assuming that $h(x,t)$ is smooth at the lattice level, the
error of discretization is ${\mathcal O}(\Delta x^{2})$. Figure
\ref{Finterface} shows the details of an interface generated using
Eq. (\ref{org_dis}). In all numerical simulations in this letter, we
put $\nu_0=D_0=1$. Other choices can be recast into this form by
rescaling the height and the time scales
\cite{Amar}. Here, we have taken $\lambda_0=3$ corresponding to medium
nonlinearity and $\Delta t=0.01$ and a smaller $\Delta t$ gives
similar results. Existence of microscopic roughness is evident and
hence errors due to the spatial discretization are uncontrolled.
Therefore, there is no $a priori$ reason why Eq. (\ref{org_dis}) should
approximate Eq. (\ref{KPZ}) unless there exist special reasons such as
conservation laws as in the linear $\lambda_0=0$ case. 

\begin{figure}
\epsfysize 8in \epsfbox{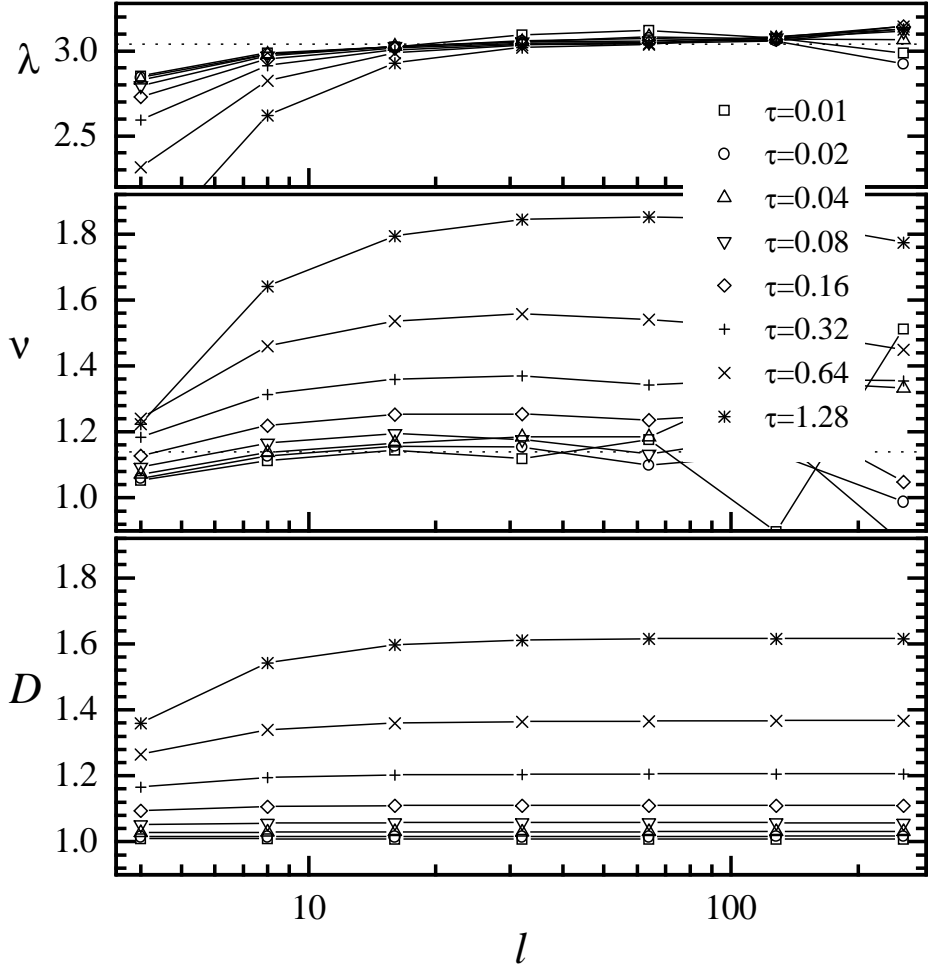}
\caption{Inverse method results on the continuum parameters $\lambda$,
$\nu$ and $D$ as functions of the spatial and temporal
resolutions $l$ and $\tau$ respectively. The dotted lines are
$\lambda=3.04$ and $\nu=1.14$.
\label{Finv}}
\end{figure}

We apply the inverse method \cite{Lam} to examine interfaces generated
using Eq. (\ref{org_dis}) with the same parameters $\nu_0=D_0=1$,
$\lambda_0=3$ and $\Delta t=0.01$ on a lattice of size $L=32768$. This
approach computes all the parameters in
the corresponding continuum KPZ description in Eq. (\ref{KPZ}). The
method extracts the continuum parameters by requiring them, when
plugged into Eq. (\ref{KPZ}), to give the best prediction on the
evolution of a surface coarse-grained up to length $l$ during a period
$\tau$. Hence $l$ and $\tau$ are respectively the spatial
and temporal resolutions of observation \cite{Lam}. Figure
\ref{Finv} shows the results. Continuum parameters are extracted at
large $l$ where finite size effects are insignificant. We obtain the
unrenormalized parameter $\lambda\simeq 3.04$. The values $\nu$ and
$D$ are renormalized in the same way due to a fluctuation dissipation
theorem \cite{Medina} and we have $D/\nu\simeq 0.88$ for all
$\tau$. At large $l$, $\tau$ controls the extent of renormalization 
since it dictates how short the wavelength of the modes should be
to evolve fast enough to contribute to renormalization
\cite{Lam}. At small $\tau$ corresponding to the short time limit in
which no renormalization has taken place, we obtain the unrenormalized
parameters $D \simeq 1.007$ and $\nu
\simeq 1.14$. We thus have $\lambda\simeq\lambda_0$ and $D\simeq D_0$
consistent with the validity of the finite difference
approximation. Unfortunately, it is clear that $\nu \neq \nu_0$.  We
check the result by calculating the ratio $D/\nu$ independently from
the correlation function \cite{Krug}:
\begin{eqnarray} 
\label{corr}
C(r)=<[h(x+r,t)-h(x,t)]^2>=
(D/\nu)
r,
\end{eqnarray} 
where the last equality is true for large $r$.  We obtain $D/\nu\simeq
0.86$ in reasonable agreement with the inverse method estimate. We
have also repeated the measurements for a much smaller $\Delta
t=0.00125$. The ratio $D/\nu$ estimated from both the inverse method
and the correlation function is indistinguishable from the previous
results within our statistical error which is less than $\pm
0.02$. Our estimates of $D/\nu$ in the range 0.86 to 0.88 are
distinctly different from $D_0/\nu_0=1$. We conclude that the discrete
equation (\ref{org_dis}) is in the KPZ universality class and is
closely related to but does $not$ approximate the continuum KPZ
equation (\ref{KPZ}). This point will be explained later.

It should be noted that decreasing $\Delta x$ is not a valid
way to improve the accuracy of the finite difference scheme but is
simply equivalent to diminishing the nonlinear parameter $\lambda$.
This is because any value of $\Delta x$ can be rescaled back to 1 by
the transformation $x
\rightarrow (\Delta x)^{-1}x$, $t \rightarrow (\Delta x)^{-2}t$, $h
\rightarrow (\Delta x)^{-1/2}h$ which leaves
Eq. (\ref{KPZ}) invariant except that $\lambda$ is now replaced by
$(\Delta x)\lambda$. In general, maintaining sufficient
nonlinearity of the system is essential to
exhibit any relevant properties of the KPZ class. It is most
convenient to fix $\Delta x=1$ and adjust the nonlinearity using
$\lambda$ as in Ref. \cite{Amar,Moser91,Review}, although tuning the
nonlinearity with $\Delta x$ has also been done \cite{Moser94}.

To further understand the anomaly, it is instructive to study
the following novel discretization which does give a correct
diffusion coefficient:
\begin{eqnarray}
\label{CEL} 
\frac{d h_i}{d t} = 
 \nu_0 \Gamma_i + \frac{\lambda_0}{2} \Psi_i + \eta_i(t),
\end{eqnarray}
for $i=1$ to $L$ with periodic boundary conditions, where
\begin{eqnarray} 
\Gamma_i&=&h_{i+1}+ h_{i-1} -2 h_{i} \label{Gamma}\\
\Psi_i&=&({1}/{3})[ ({h_{i+1}-h_{i}})^2 
+({h_{i+1}-h_{i}}) ({h_{i}-h_{i-1}}) \label{Psi}
\nonumber\\& & 
+({h_{i}-h_{i-1}})^2 ].
\end{eqnarray}
The noise $\eta_i(t)$ has a Gaussian distribution and
$<\eta_i(t)\eta_j(t')> = 2 D_0 \delta_{ij} \delta(t-t')$.  Both
Eqs. (\ref{org_dis}) and (\ref{CEL}) could be equally valid ${\mathcal
O}(\Delta x^{2})$ spatial discretizations of Eq. (\ref{KPZ}) if the
interface were smooth. However, neither are necessarily a good finite
difference approximation of Eq. (\ref{KPZ}) due to the microscopic
roughness. 

The steady state properties of the new equations admit elegant
exact solutions. Let $P[h,t]$ be the probability distribution of the
discrete interface $\{h_i\}_{i=1}^L$ at time $t$. It obeys the
Fokker-Planck equation \cite{Review}:
\begin{eqnarray} 
\label{Fokker}
\frac{\partial P}{\partial t} = 
- \sum_{i=1}^{L} \frac{\partial}{\partial h_i} 
 \left[ \left( \nu_0 \Gamma_i + \frac{\lambda_0}{2} \Psi_i \right)
 P \right] 
 + D_0  \sum_{i=1}^{L} \frac{\partial^2 P}{\partial h_i^2 } 
\end{eqnarray}
which has the exact steady state solution:
\begin{eqnarray} 
\label{P}
P[h]=\exp\left[-\frac{\nu_0}{2D_0} \sum_{i} (h_{i+1}-h_i)^2 \right] .
\end{eqnarray}
The existence of this exact solution results from the vanishing of the
$\lambda_0$ term on the RHS of Eq. (\ref{Fokker}) when Eq. (\ref{P})
is applied, in complete
analogy with the continuum case \cite{Review}. The specific form of
$\Psi_i$ in Eq. (\ref{Psi}) is specifically chosen to allow for the
cancelation and this property is not shared by other discretizations
in general.

	 We now calculate the associated continuum
parameters in the KPZ description of Eq. (\ref{CEL}). Assuming a large
lattice, it follows from Eq. (\ref{P}) that the correlation function
is $C(r)=(D_0/\nu_0)r$. Comparing with the continuum result in
Eq. (\ref{corr}), we obtain $D/\nu=D_0/\nu_0$. At the short time
limit, the noise terms dominate in both Eqs. (\ref{KPZ}) and
(\ref{CEL}) and it is easy to see that the continuum short time noise
parameter is $D=D_0$. Hence, the short time continuum diffusion
coefficient is $\nu=\nu_0$.  To calculate $\lambda$, we consider a
screw boundary condition so that the interface has an average slope
$u$. The steady state probability distribution now becomes $
P[h]=\exp\left[-({\nu_0}/{2D_0}) \sum_{i} (h_{i+1}-h_i-u)^2 \right]
$. It is then easy to show that the average growth
velocity is $
v(u)=\left<{\partial h_i}/{\partial t}\right>
 = {\lambda_0}/{3} + {\lambda_0}{}u^2/2 
$. The continuum nonlinear parameter can be calculated from $\lambda=
v_\infty''(0)$ \cite{Krug} and we get $\lambda=\lambda_0$. The average
growth velocity is $c=v(0)=\lambda_0/3$. Therefore, all three
continuum parameters $\nu$, $\lambda$ and $D$ are exactly the
respective nominal values $\nu_0$, $\lambda_0$ and $D_0$ in the
new discretization. These results are confirmed numerically using both
the inverse method and measurements of correlation functions.

	To gain further insights, we calculate the short time value of
$\nu$ directly by a novel analytical application of the inverse
method. When calculating the continuum parameters $c$, $\nu$ and
$\lambda$ disregarding any higher order terms, the inverse method
reduces the problem to the solution of a matrix equation
\cite{Lam}. Due to an up-down symmetry of the interfaces at long
length scales \cite{Medina}, the matrix is block
diagonal at large spatial resolution $l$ and the expression for $\nu$ is simplified to
\begin{eqnarray}
\label{nu}
\nu=\frac{<(\partial h/\partial t)_c (\partial^2 h/\partial x^2)_c>}
{<(\partial^2 h/\partial x^2)_c^2>}
\end{eqnarray}
where the subscript $c$ denotes coarse graining to length scale $l$
before evaluation at a given lattice point $i$. The operation
$\partial /\partial x$ is usually carried out in the Fourier space. The
growth rate $(\partial h/\partial t)_c$ is contributed by $\Gamma$
and $\Psi$ in Eqs. (\ref{CEL})-(\ref{Psi}) while effects of the noise
vanish after averaging. Now the steady state distribution $P[h]$ in
Eq. (\ref{P}) is invariant under the symmetry operation $h\rightarrow
-h$, while the nonlinear term $\Psi$ has the opposite parity from
that of $\partial^2 h/\partial x^2$. Therefore, $<\Psi_c (\partial^2
h/\partial x^2)_c>=0$. In contrast, one can easily show that the
remaining linear discrete diffusion term $\Gamma$ approaches
$(\partial^2 h/\partial x^2)_c$ at sufficient coarse graining. Hence
we get from Eq. (\ref{nu}) that $\nu=\nu_0$ confirming our previous
arguments.

We now re-examine the conventional discretization in
Eq. (\ref{org_dis}) in light of our new results.  In this case, the
interface distribution $P[h]$ has no simple solution in general. An
exception is the linear $\lambda_0=0$ case in which
Eqs. (\ref{org_dis}) and (\ref{CEL}) become identical. Then
Eq. (\ref{P}) is again the exact steady state solution and we have
$\nu=\nu_0$, $D=D_0$ and $\lambda=\lambda_0=0$.  A finite $\lambda_0$
perturbs the system. We found numerically at $\lambda_0=3$ and
$\nu_0=D_0=1$ that the interface distribution $P[h]$ is not far from
that in Eq. (\ref{P}). However, there is a small skewed correlation
among the height differences of neighboring lattice points which can be
exemplified by a skewness in the probability distribution of
$\Gamma_i$ defined in Eq.  (\ref{Gamma}). As a result, the up-down
symmetry of $P[h]$ is broken and Eq. (\ref{nu}) now gives
$\nu=\nu_0+<\Psi_c (\partial^2 h/\partial x^2)_c> /<(\partial^2
h/\partial x^2)_c^2>\neq\nu_0$. The proof of $D=D_0$ is similar to the
previous case.  Our numerical results favor a slightly larger
$\lambda$ than $\lambda_0$ instead of an equality.
It can be caused by some non-trivial dependence of $P[h]$ on the
inclination in contrast to that for the new discretization.

The new discretization is also a valuable tool for numerical
investigations. Realizations of steady state interfaces could be
generated directly using the exact distribution in Eq. (\ref{P}). To
simulate the dynamics, the equations can be integrated
using Euler's method which has an error ${\mathcal O}(\Delta t^{1/2})$
for stochastic equations \cite{Kloeden}.
However, an operator splitting approach is far more efficient. Each
iteration now consists of two half steps:
\begin{eqnarray} 
h^{n+\frac{1}{2}}&=&{\mathcal U}_L[h^n,\Delta t]
\label{UL} \\
h^{n+1}&=&{\mathcal U}_N[h^{n+\frac{1}{2}},\Delta t]
\label{UN}
\end{eqnarray}
where ${\mathcal U}_L$ acting on $\{h^n_i\}_{i=1}^L$ is the stochastic
linear evolution operator for $\Gamma$ and $\eta$ only in
Eq. (\ref{CEL}), while ${\mathcal U}_N$ is the nonlinear evolution
operator for the remaining $\Psi$ term. By noting that $P[h]$ in
Eq. (\ref{P}) is also the steady state solution of both ${\mathcal
U}_L$ and ${\mathcal U}_N$ $independently$, it is easy to show that
the continuum parameters $c$ and $\lambda$ and the short time
parameters $\nu$ and $D$ derived above remain $exact$. Hence the
relevant dynamics is not perturbed despite an ${\mathcal O}(\Delta
t^{1/2})$ error in the particular realization of the interface. The
linear operator ${\mathcal U}_L$ can be handled exactly and
Eq. (\ref{UL}) implies, after some algebra,
\begin{eqnarray} 
\label{linear}
h^{n+\frac{1}{2}}_i=\sum_{j=1}^L K^\Gamma_{i-j} h^n_j 
 + \sqrt{2D_0\Delta t}\sum_{j=1}^L K^\eta_{i-j} \xi^n_j 
\end{eqnarray}
where the $\xi^n_j$'s are independent standard Gaussian variables.
The propagators $K^\Gamma$ and $K^\eta$ are computed from their
 Fourier coefficients:
\begin{eqnarray} 
\tilde{K}^\Gamma_k&=&\exp(-\gamma_k \Delta t)\\
\tilde{K}^\eta_k&=&\left\{ \begin{array}{ll}
	1 & \mbox{if $k=0$}\\
	\left\{\left[ 1-\exp(-2\gamma_k \Delta t) \right]/
	(2\gamma_k \Delta t)\right\}^{1/2}
	&\mbox{if $k\neq 0$} 
	\end{array}
\right.
\end{eqnarray}
where $\gamma_k=\cos( 2\pi k/L)$ and $k$ is an integer from $0$ to $L-1$. In practice, the propagators are sharply
peaked so that summation over only $\mid i-j \mid
\leq 5$ in Eq. (\ref{linear}) is sufficient. The deterministic
evolution in Eq. (\ref{UN}) can be integrated using the fourth-order
Runge-Kutta method with an error ${\mathcal O}(\Delta t^4)$
\cite{Press} which is
also the overall error of our approach.

\begin{figure}
\epsfysize 8in \epsfbox{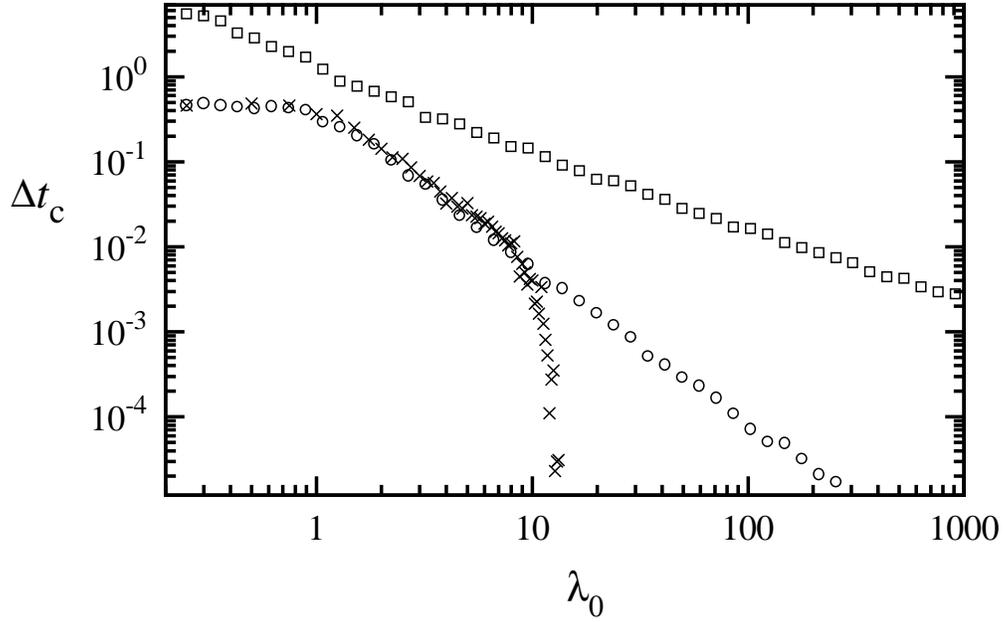}
\caption{Largest possible time step $\Delta t_c$ for numerical
stability against $\lambda_0$ for Eq. (\ref{CEL}) integrated respectively by
operator splitting approach ($\Box$) and Euler's method ($\circ$), and for
Eq. (\ref{org_dis}) integrated by Euler's method ($\times$).
\label{Fstab}}
\end{figure}

	To test the numerical stability, we simulate initially flat
interfaces of size $L=128$ at $\nu_0=D_0=1$ with various values of
$\lambda_0$ each for a period $10000/\lambda_0$. Figure \ref{Fstab}
plots $\Delta t_c$ against $\lambda_0$ where $\Delta t_c$ is the
critical time step just small enough to ensure stability during a run.
For the new discrete equations  integrated using the operator splitting
Runge-Kutta approach and Euler's method, we found respectively $\Delta
t_c\sim \lambda_0^{-0.89}$ and $\Delta t_c\sim \lambda_0^{-1.76}$ for
$\lambda_0\agt 1$. For the conventional approach of Euler's algorithm
in Eq. (\ref{org_dis}), the result is initially similar to that of the
same method applied to the new discretization. However, $\Delta t_c$
drops faster than exponentially at $\lambda_0 \agt 8$. We suspect that
there is a critical $\lambda_0$ beyond which instability is
unconditional.  For Euler's method, $\Delta t_c=0.5$ at $\lambda_0\alt
1$ because of an instability of the diffusive term
\cite{Press}.  The above findings should be important for
further understanding of numerical instabilities in equations for
interfaces. The superior performance of our discretized equations
when integrated with the operator splitting method is evident. The
exponent $-0.89$ in fact represents a surprisingly good stability,
taking into account that the dynamic time scale of the interface is
proportional to $\lambda_0^{-1}$ for large $\lambda_0$.  Besides
improved stability, it is about 30 times faster than the conventional
approach for example at $\lambda_0=3$ for a small systematic
error of less than $0.05\%$ in the average growth velocity $c$.  

The KPZ equation can be recast using a Hope-Cole transformation
into a form which describes directed polymers in random media. It has
been reported in Refs. \cite{Beccaria} and \cite{Newman} that the
discretized versions of the transformed equation are numerically more
stable than those obtained directly from the KPZ equation. Yet, our
numerical work shows that their stability and hence the computational
efficiency is only in between those of the conventional and our new
discrete equations in 1+1 dimensions. These discretizations also
suffer from the shortcoming that the effective diffusion coefficient
is incompatible with the nominal value. Nevertheless, these methods
work also in higher dimensions while it is not clear how our
discretization can be generalized.

In conclusion, we have explained that the conventional finite
difference approach does $not$ provide a genuine direct numerical
integration of the KPZ equation since the continuum diffusion
coefficient is incompatible with the nominal one in the discrete
equations. This is explained by the microscopic roughness and skewness
of the interface at steady state. Despite this anomaly, the
discrete equations themselves are in the KPZ universality class and
thus the scaling exponents measured in previous works are all valid.
A novel discretization for the KPZ equation is studied. The continuum
diffusion coefficient is hence shown analytically to be equal to the
nominal value. However, this equality is due to subtle
cancelation of terms in the Fokker-Planck equation and should not
hold in general. This letter has focused on the KPZ equation in $1+1$
dimensions.  However, the conventional finite difference integration
approach is also routinely applied to growth in higher dimensions as
well as to variants of the KPZ equation and related problems such as
the Kuramato-Sivaskinsky equation, etc. \cite{Review}. It should be
interesting to examine limitations of those results in light of our
findings.  There are few investigations on the generalization of
the KPZ equation with higher order terms \cite{Amar93}, although they
may be necessary for successful renormalization group calculations
\cite{Drossel}. Our refined understanding on the integration approach
should be important for generating such higher order terms in a
controlled manner.

	We thank L.M. Sander and J. Li for interesting
communications. This work is supported by RGC Grant No.0354-046-A3-110
and PolyU Grant No. 0353-003-A3-110.

\newpage

\newpage

\end{document}